\documentclass[12pt]{article}
\usepackage{latexsym,graphicx,multirow}
\usepackage{amssymb}
\usepackage{amsmath}
\usepackage{amscd}
\usepackage{amsthm}
\usepackage{float}

\usepackage[left=2cm,top=2.5cm,right=2.5cm,bottom=1.5cm]{geometry}
\usepackage{graphicx}
\usepackage{epstopdf}
\usepackage{graphicx,epstopdf}
\usepackage{hyperref}
\DeclareGraphicsExtensions{.eps}
\usepackage{epstopdf} 

\begin{document}

	\begin{center}
		\large{\bf
		Tsallis	holographic model of dark energy: Cosmic behaviour, statefinder analysis and $\omega_{D}-\omega_{D}^{'}$ pair in the non- flat universe} \\
		\vspace{10mm}
		\normalsize{ Vipin Chandra Dubey$^1$, Umesh Kumar Sharma$^2$, A. Beesham $^3$  }\\
		\vspace{5mm}
		\normalsize{$^{1,2}$Department of Mathematics, Institute of Applied Sciences and Humanities, GLA University\\
			Mathura-281 406, Uttar Pradesh, India}\\
		$^{3}$Department of Mathematical Sciences, University of Zululand, Kwa-Dlangezwa 3886, South Africa\\
		\vspace{2mm}
		$^1$E-mail: vipin.dubey@gla.ac.in.\\
         $^2$E-mail:  sharma.umesh@gla.ac.in\\
         $^3$E-mail: beeshama@unizulu.ac.za\\
			\vspace{2mm}
		\vspace{5mm}
		\vspace{10mm}
	
\end{center}
\begin{abstract}
 The paper investigates the Tsallis holographic dark energy (THDE) model in accordance with the apparent horizon as an infrared cut - off, in a non- flat universe. The cosmological evolution of the deceleration parameter and equation of state of THDE model are calculated. The evolutionary trajectories are plotted for the THDE model for distinct values of the Tsallis parameter $\delta$ besides distinct
spatial curvature contributions, in the statefinder $(r, s)$ parameter-pairs and $\omega_{D}-\omega^{'}_{D}$ plane, considering the present value of dark energy density parameter
$\Omega_{D0} $, $\Omega_{D0}=0.72$, in the light of $WMAP + eCMB + BAO + H_{0}$ observational data.
The statefinder and $\omega_{D}-\omega^{'}_{D}$ plane plots
specify the feature of the THDE and demonstrate the separation
between this framework and other models of dark energy.
\end{abstract} 
\section{Introduction}
The dark area of the cosmos intrigues the scholars and the latest surveys
\cite{ref1,ref2,ref3,ref4,ref5,ref6,ref7,ref8,ref9,ref10}, demonstrate that the dark sector contribution is in roughly $95\%$ in comparison to substratum ;
rest is negligible radiation and baryonic matter (about $4 - 5\%$). The baryonic matter comprises electromagnetic
radiation
and may be seen precisely \cite{ref11}. The dark matter, which around $25\%$ of the
matter part of the universe is a pressure less matter, which gives
the velocity of galaxy clusters \cite{ref12}. It was inferred that there was more mass in galaxies; than observed by the corroboration CDM with the examinations of rotational trajectories of spiral galaxies \cite{ref13}. This exotic matter presence is affirmed by the gravitational lensing and galaxy clusters on the basis of x-ray emission.\\

The dark energy, around $70\%$ of the rest cosmic foundation, is responsible for the current accelerated stage of the universe \cite{ref14,ref15}. Under the standard cosmological description, the cosmological constant $\Lambda$ may be used for DE component, which is having a geometric nature in accordance with the GR with constant energy density and EoS (equation of state) $\omega_{D} = -1$ \cite{ref16,ref17}. The description of dark energy is in toe with
recent observational data, but it opposes the theoretical
prediction for vacuum energy, which has been observed by quantum field theory \cite{ref18}. This material description has the inflationary paradigm,
the Big Bang nucleosynthesis (BBN) and general relativity (GR), called the $\Lambda$CDM model. Normally dynamical description is taken for DE part via a
distinct EoS, instead of the vacuum description, for example, EoS parameter $\omega \neq - 1$ (constant or time-dependent) \cite{ref19,ref20}. Other proposed DE component are the
dynamical approach through a scalar field \cite{ref21,ref22} and modified theories of
gravity \cite{ref23,ref24,ref25,ref26}.\\

Holographic dark enegy (HDE) is an interesting attempt to explain the current accelerated expansion of the universe \cite{ref27,ref28,ref29,ref30,ref31,ref32,ref33,ref34,ref35,ref36,ref37,ref38,ref39,ref40,ref41,ref42,ref43,ref44} in line with some observations \cite{ref45,ref46,ref47,ref48,ref49}, within the framework of holographic rule \cite{ref51,ref52,ref53} by characterizing and portraying the established holographic energy density as $\rho_{D}= 3c^{2}M^{2}_{pl}L^{-2}$, contingent upon the entropy connection of black holes \cite{ref27}, here c denotes constant. Such models are in incredible concurrence with ongoing observations and have drawn the consideration of scientists as of late. Also, a few
other holographic rule persuaded DE models have likewise been recommended \cite{ref54,ref55,ref56}.\\

Recently, various DE models being produced for deciphering or portraying the accelerated expansion of the universe, are accessible. With a specific genuine goal to be able to confine between these battling cosmological scenarios including DE, a delicate and exceptional feature for DE models is an absolute need. Thus, a logical suggestion that impacts the utilization of the determination of pair $(r, s)$, the specified "statefinder", was suggested by Sahni et al. \cite{ref57,ref58}. It is not elusive that the statefinder parameters to refine the universe evolution components through the higher-order derivative of the scale factor (a) and is a trademark following stage past Hubble parameter ($H$) and deceleration parameter ($q$). Since various cosmological models including DE demonstrate dynamically special advancement directions in the $s-r$ plane, consequently, statefinder is a helpful tool for seeing these DE models which are characterized as \\

\begin{eqnarray}
\label{eq1}
r = \frac{\dddot {a}}{aH^{3}}, \hspace{1cm} s = \frac{r-1}{3 (q - \frac{1}{2})}.
\end{eqnarray}

The geometrical analysis, statefinder is constructed directly from space-time metric, as it is considered as universal, is basically model-dependent in describing the dark energy. To find the qualitatively different behaviours to that of models of dark energy, the $r-s$ plane is taken into consideration for plotting the evolutionary trajectories. The spatially flat $\Lambda$CDM is defined by the fixed point i.e.
$({r, s}) = ({1, 0})$. The displacement of the derived model from $\Lambda$CDM is calculated by the displacement from this fixed point to a dark energy model, which provides a good way of observing
this distance. As explored in
\cite{ref58,ref59,ref60,ref61,ref62,ref63,ref63a,ref63b,ref63c,ref63d,ref64,ref65,ref66,ref67,ref68,ref69} the statefinder may effectively differentiate between
a wide variety of dark energy models including quintessence, the cosmological
constant, braneworld and the Chaplygin gas models and
interacting dark energy models. Recently, Sharma
and Pradhan (2019) and Gunjan et al. (2019) have diagnosed
the geometric behaviour of non-interacting and
interacting THDE for a flat universe in terms of statefinder parameters and
$\omega_{D} - \omega_{D}^{'}$ pair in detail \cite{ref70,ref71}. Sheykhi (2018), have derived the modified Friedmann equations for an FRW universe with the apparent horizon as IR cutoff in the form of Tsallis entropy \cite{ref86a}. \\

In this work, we apply the statefinder diagnosis and $\omega_{D} - \omega_{D}^{'}$ plane diagnostic to the THDE model
in the non-flat universe with the apparent horizon as IR cutoff. The statefinder may likewise be utilized to analyze distinctive illustration of the model, including different model parameters and distinctive spatial curvature ($\Omega_{K0}$) inputs.
Moreover, we consider the values of
$\Omega_{K0} =$ $-0.0027$, $0.001$ and $0$
corresponding to the open, closed and flat universes, respectively, in the light of $WMAP + eCMB + BAO + H_{0}$ \cite{ref6} and by
varying the Tsallis parameter $\delta$ \cite{ref75,ref76}.\\

The sequence of the paper is as follows: In Sec. 2, we review the
THDE considering the IR cutoff as the apparent horizon. In Sec. 3, observational data is given. In Sec. 4, the cosmic behaviour of the deceleration parameter and EoS is described. In Sec. 5,
the statefinder parameters are obtained and plotted. In Subsec. 5.1, $\omega_{D} - \omega_{D}^{'}$ plane is explored. Discussions and Conclusions are talked about in Sec. 6\\

\section{ THDE : Brief review with IR cutoff as apparent horizon}

 The foundation of the HDE approach is the definition of the system boundary, and
 in fact, modification in the HDE models can be done by changing the system entropy \cite{ref43,ref44,ref56,ref74}. Additionally, statistical generalized mechanics produces a new entropy for black holes \cite{ref75},
 an entropy which varies from the Bekenstein entropy and prompts another HDE model called THDE \cite{ref76}. In \cite{ref76}, THDE has been investigated with Hubble horizon as IR cutoff which corresponds to the accelerated universe but was not stable. Recently, the THDE is investigated in different scenario with other IR cutoffs \cite{ref77,ref78,ref79,ref80,ref81,ref82,ref83,ref84}. Here, we have considered, the apparent horizon as IR cutoff as the usual system boundary for the FRW universe situated at \cite{ref86a,ref86,ref87,ref88}.
 \begin{eqnarray}
 \label{eq2}
 \tilde{r}_A=\frac{1}{\sqrt{\frac{k}{a^2}+H^2}}
 \end{eqnarray}
 Gibbs in 1902, pointed out that the standard Boltzmann-Gibbs theory is not suitable for divergent partition functions
 and large-scale gravitational systems are in line this class. Also, it has been contended that frameworks including the long-range interactions may fulfil statistical generalized mechanics rather than the Boltzmann-Gibbs insights \cite{ref81}. Tsallis \cite{ref75} standard passes Gibbs theory as a limit. So, additive Boltzmann–Gibbs entropy must be generalized to the
 nonextensive, i.e non-additive entropy, which is called Tsallis entropy.
 The black hole horizon entropy can be
 modified as \cite{ref75}
 \begin{eqnarray}
 \label{eq3}
 S_{\delta }=\gamma A^{\delta }
 \end{eqnarray}
 where $\delta$ represents the non-additive parameter and $\gamma $ denotes unknown constant. Since a null hypersurface is represented by $\tilde{r}_A$ for the FRW spacetime
 and also, a proper system boundary
 \cite{ref86,ref87,ref88}, a property equivalent to that of the black hole horizon, one may observe
 $\tilde{r}_A$ as the IR cutoff, and utilize the holographic DE speculation to get \cite{ref81,ref84}.
 \begin{eqnarray}
 \label{eq4}
 \rho _D=B\tilde{r}_A{}^{2 \delta -4}
 \end{eqnarray}
 where B denotes the unknown parameter. It was contended that the entropy related with the
 apparent horizon of FRW universe, in every gravity hypothesis, has a similar structure as the same structure as the entropy of black hole horizon in the relating gravity. The
 just change one need is supplanting the black hole horizon span $ r_+ $ by the apparent
 horizon radius $\tilde{r}_A$ \cite{ref81}.\\
 
 The metric for FRW non-flat universe is defined as :
 \begin{eqnarray}
 \label{eq5}
 ds^{2} = -dt^{2}+a^{2}(t)\Big(\frac{dr^{2}}{1-kr^{2}}+ + r^{2}d\Omega^{2}\Big)
 \end{eqnarray}
 where $ k = -1$, $ 1 $ and $0$ represent a open, closed and flat universes, respectively.
 The first Friedmann equation in a non-flat FRW universe, including THDE and darkmatter (DM) is given as :
 
 \begin{eqnarray}
 \label{eq6}
 H^2 + \frac{k}{a^2}=\frac{1}{\tilde{r}^2{}_A}=\frac{1}{3} (8 \pi G) \left(\rho_D+\rho _m\right),
 \end{eqnarray}

 where $\rho_{m}$ and $\rho_{D}$ represent the energy density of matter and THDE, respectively, and $\frac{\rho_{m}}{\rho_{D}} = r$ represent the ratio of energy densities of two dark sectors \cite{ref43,ref81}. Using the fractional energy densities, the energy density parameter of pressureless matter, THDE and curvature term can be expressed as \\
 
 \hspace{2cm} $\Omega_{m} = \frac{8\pi\rho_{m}G}{3H^{2}} $, \hspace{1cm} $\Omega_{D} = \frac{8\pi\rho_{D}G}{3H^{2}} $ \hspace{1cm} $\Omega_{k} = \frac{k}{a^{2}H^{2}}$\\

 Now Eq. (\ref{eq6}) can be written as:
 \begin{eqnarray}
 \label{eq7}
 1 +\Omega _k = \Omega _D+\Omega _m
 \end{eqnarray}
 The law of conservation for matter THDE and are given as :
 \begin{eqnarray}
 \label{eq8}
 \dot \rho_{m} + 3 H \rho_{m} = 0
 \end{eqnarray}
 \begin{eqnarray}
 \label{eq9}
 \dot \rho_{D} + 3H (\rho_{D} + p_{D}) = 0
 \end{eqnarray}
 in which $ \omega _D = p _D/\rho _D$ represents the THDE EoS parameter.
 Combining with the definition of $r$, we get

 \begin{eqnarray}
 \label{eq10}
 r=\frac{\Omega _K+1}{\Omega _D}-1
 \end{eqnarray}
 
 Now, using derivative with time of Eq. (\ref{eq6}) in Eq. (\ref{eq9}), and combined the result with
 Eqs. (\ref{eq8}) and Eq. (\ref{eq7}), we get

 \begin{eqnarray}
 \label{eq11}
 \frac{\dot{H}}{H^2}=\Omega _K-\frac{3}{2} \Omega _D \left(\omega _D+r+1\right)
 \end{eqnarray}
 
 Using Eq. (\ref{eq11}), The deceleration parameter $q$ is obtained as
 \begin{eqnarray}
 \label{eq12}
 q=-\frac{\dot{H}}{H^2}-1=\frac{3}{2} \Omega _D \left(\omega _D+r+1\right)-\Omega _K-1
 \end{eqnarray}
 Now, using the derivative with time of Eq. (\ref{eq4}) with Eqs. (\ref{eq2})
 and Eq. (\ref{eq11}), we get
 \begin{eqnarray}
 \label{eq13}
 \dot{\rho _D}=\frac{3 (\delta -2) H \rho _D \Omega _D \left(\omega _D+r+1\right)}{\Omega _K+1}
 \end{eqnarray}
 Now by using the time derivative of energy density parameter $\Omega _D $ with Eqs. (\ref{eq13}) and (\ref{eq11}), one gets
 
 \begin{eqnarray}
 \label{eq14}
 \Omega _D'=\Omega _D \left(\frac{3 \Omega _D \left(\omega _D+r+1\right) \left(\delta +\Omega _K-1\right)}{\Omega _K+1}-2 \Omega _K\right)
 \end{eqnarray}
 where, dot denotes derivative with respect to time, and prime denotes the derivative with respect to the ln a.\\
 
 Additionally, calculations for the density parameter
 
 \begin{eqnarray}
 \label{eq17}
 \Omega '_{D}=\frac{\Omega _D \left(\Omega _D \left(-3 \delta +(1-2 \delta ) \Omega _K+3\right)+\left(\Omega _K+1\right) \left(3 \delta +\Omega _K-3\right)\right)}{(\delta -2) \Omega _D+\Omega _K+1}
 \end{eqnarray}

\section{\bf Observational Data}

In this section, we introduce the observational data used in the analysis of Tsallis holographic dark energy model.\\

The streamlining presumptions of cosmic scale homogeneity and isotropy guide to the well-known
FLRW metric that has all the earmarks of being an exact depiction of our universe. The base $\Lambda$CDM cosmology expect a FLRW
metric with a flat 3-space. This is a prohibitive presumption
that should be tried exactly. For FLRW models
$K > 0$ compares to adversely bended 3-geometries while
$K < 0$ compares to emphatically bended 3-geometries. Indeed, even with flawless information inside our past lightcone, our derivation of the curvature
$K$ is restricted by the cosmic variance of curvature perturbations
that are still super-horizon at the present since these can not be
recognized from background curvature inside our observable
volume.\\

The curvature parameter $K$ increases as a power law during the matter dominated and radiation era, but decreases exponentially with time during inflation, So the inflationary forecast has been that curvature ought to be imperceptibly small at present. By, adjusting parameters it is conceivable to
devise inflationary models that produce open \cite{ref91a,ref91b} or closed universes \cite{ref91c}. Indeed all the more theoretically, there has been intrigue as of late in multiverse models, in which topologically open "pocket universes"
structure by bubble nucleation \cite{ref91d,ref91e} between various vacua of a "string landscape" \cite{ref91f,ref91g}. Plainly, the recognition of a noteworthy deviation from $\Omega_{K} = 0$ would have significant ramifications for expansion hypothesis and Physics \cite{ref7}. With primary CMB fluctuations alone \cite{ref95,ref96}, the spatial curvature, $\Omega_{k}$, is constrained by the geometric degeneracy. Be that as it may, with the ongoing recognition of CMB lensing in the high-$l$ power spectrum \cite{ref97,ref98}, the degeneracy between
$\Omega_{\Lambda}$ and $\Omega_{m}$ is currently considerably decreased. This creates a huge identification of DE and tight limitations on spatial curvature utilizing just CMB information: when the ACT and SPT information, including the lensing limitations, are joined with nine-year WMAP information.
Bennett et.al \cite{ref99}, demonstrates the joint limitations on ($\Omega_{m}, \Omega_{\Lambda}$) (and $\Omega_{k}$, verifiably) from the as of now
accessible CMB information. Consolidating the CMB information with lower-redshift separate markers, such $H_{0}$, BAO, or supernovae further constrains $\Omega_{k}$ (\cite{ref90}). Expecting the DE is vacuum energy ($w=-1$), it was given \\

\hspace{3cm} $\Omega_{k} = -0.0027^{+0.0039}_{-0.0038}$ \qquad WMAP+eCMB+BAO+$H_{0}$.\\

The summary of the non-flat cosmological parameters used in this work is given in Table 1.

		\begin{table}
		\caption{\small Non-flat $\Lambda$CDM Cosmological parameters.}
		\begin{center}
			\begin{tabular}{ |c |c |}
				\hline
					Parameter  &	 WMAP + eCMB + BAO + $H_{0}$ \\
					
			\hline				
				$\Omega_{k}$	&   $ -0.0027^{+0.0039}_{-0.0038}$	 \\
			
					\hline				
						$\Omega_{tot}$	&  $1.0027^{+0.0038}_{-0.0039}$ 	 \\
				
				\hline				
						$\Omega_{m}$	&  $0.2855^{+0.0096}_{-0.0097}$	 \\
				\hline				
				$\Omega_{\Lambda}$	&   $0.717\pm 0.011$  \\
					\hline				
				$H_{0}(km/s/Mpc)$	&  $68.92^{+0.94}_{-0.95}$ \\
			
				\hline
			\end{tabular}
		\end{center}
	\end{table}

	\begin{figure}
	\begin{center}
		\includegraphics[width=16cm,height=7cm, angle=0]{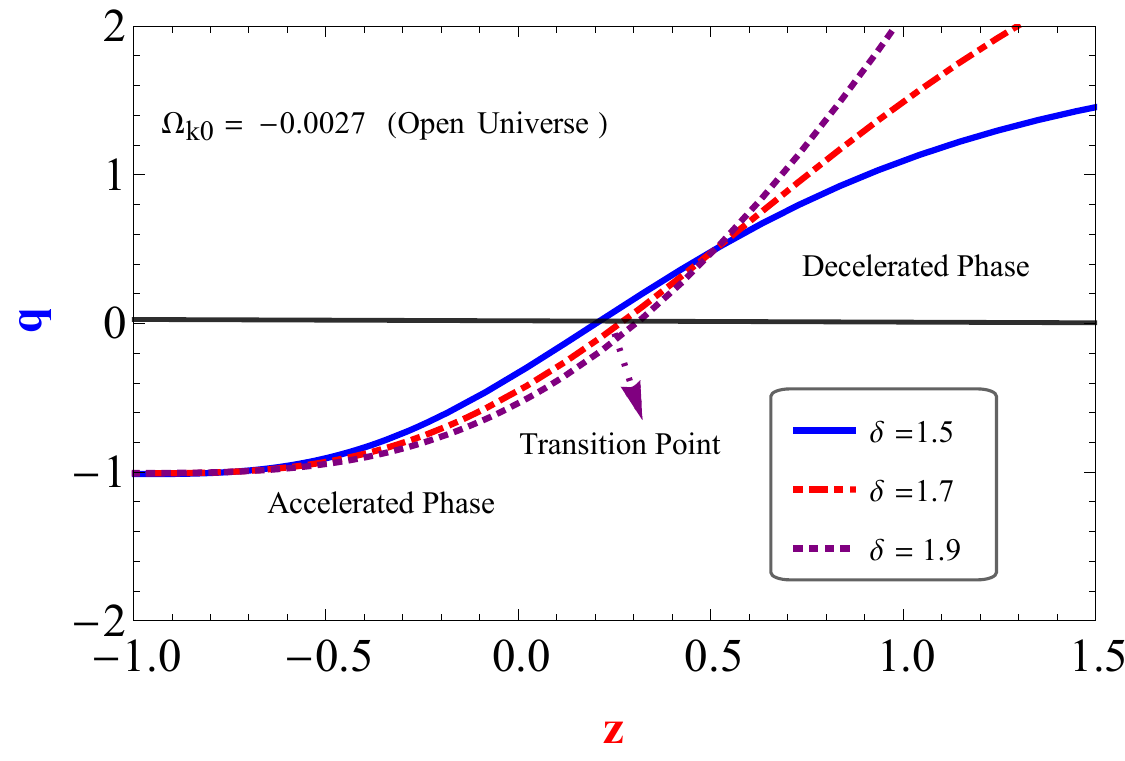}
		\includegraphics[width=16cm,height=7cm, angle=0]{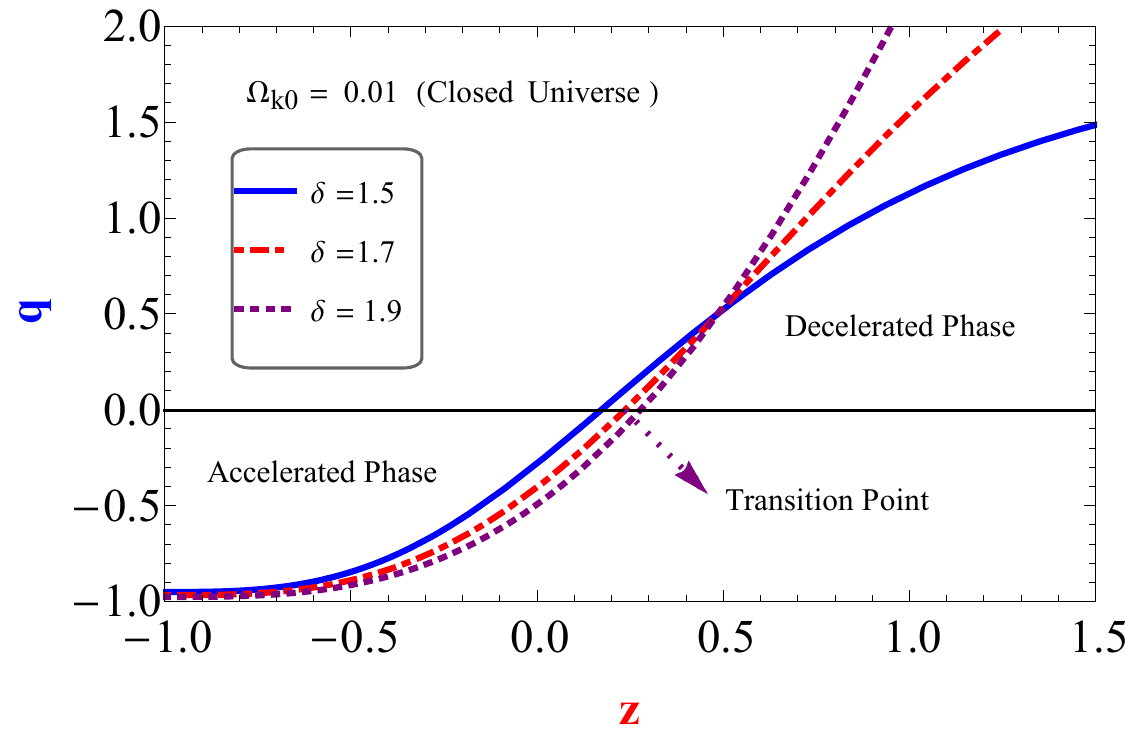}
		\includegraphics[width=16cm,height=7cm, angle=0]{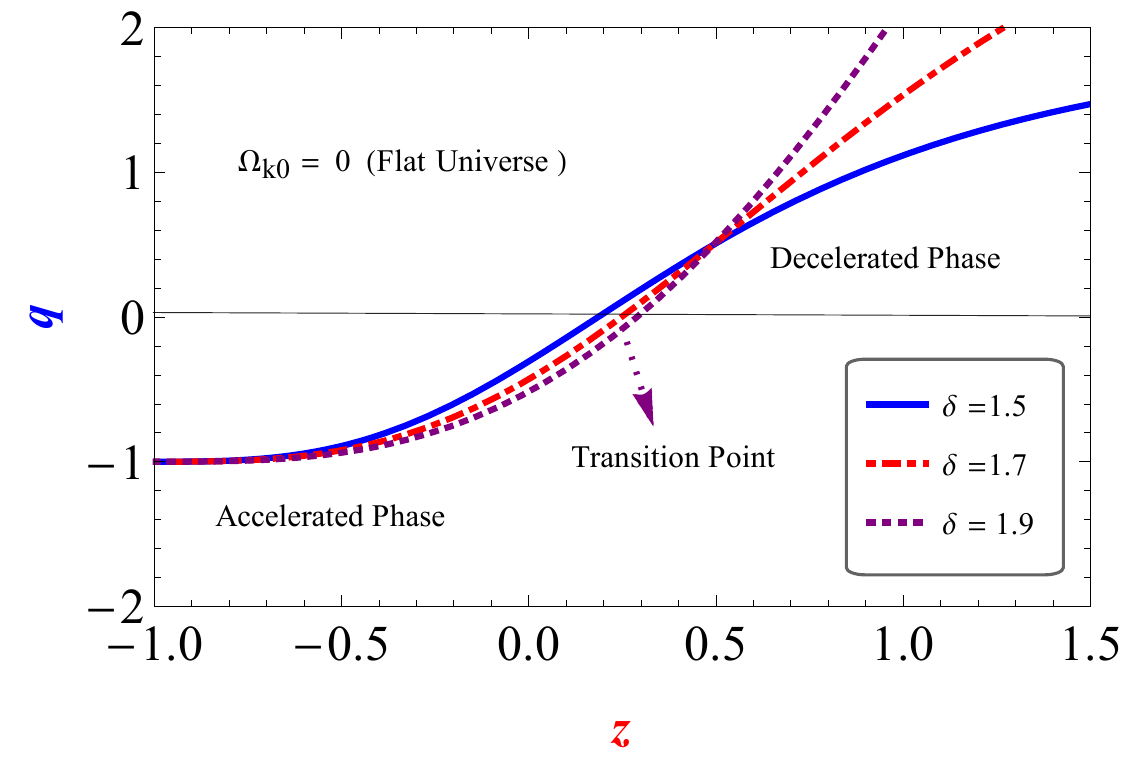}
		\caption { The evolutionary behaviour  of  deceleration parameter  ($q$)  in THDE model against redshift $z$ for different cases of Tsallis parameter $\delta = 1.5$, $\delta = 1.7$ and $\delta = 1.9$. Selected graphs are plotted for $\Omega_{D0}= 0.72$ and taking $\Omega_{K0}$= $ -0.0027$, $0.001$ and  $0$  corresponding to open, closed and flat  universes, respectively, in the light of $WMAP + eCMB + BAO + H_{0}$ observational data.} 
		
	\end{center}
\end{figure}

\begin{figure}
	\begin{center}
		\includegraphics[width=16cm,height=7cm, angle=0]{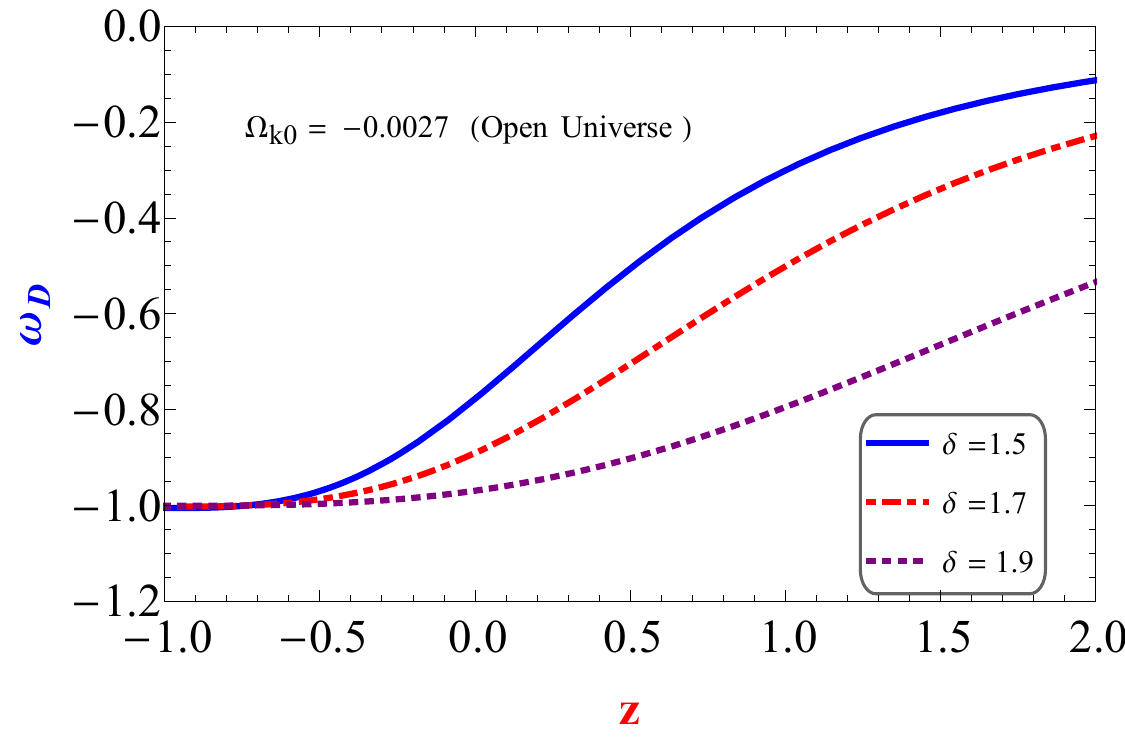}
		\includegraphics[width=16cm,height=7cm, angle=0]{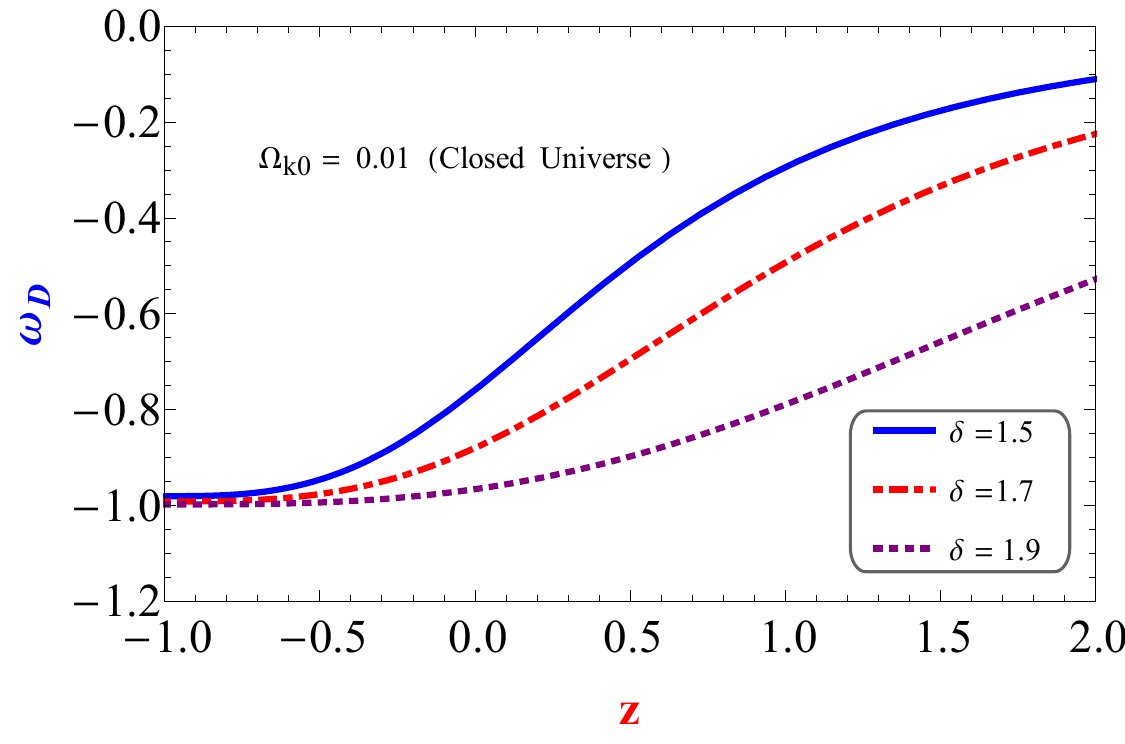}
		\includegraphics[width=16cm,height=7cm, angle=0]{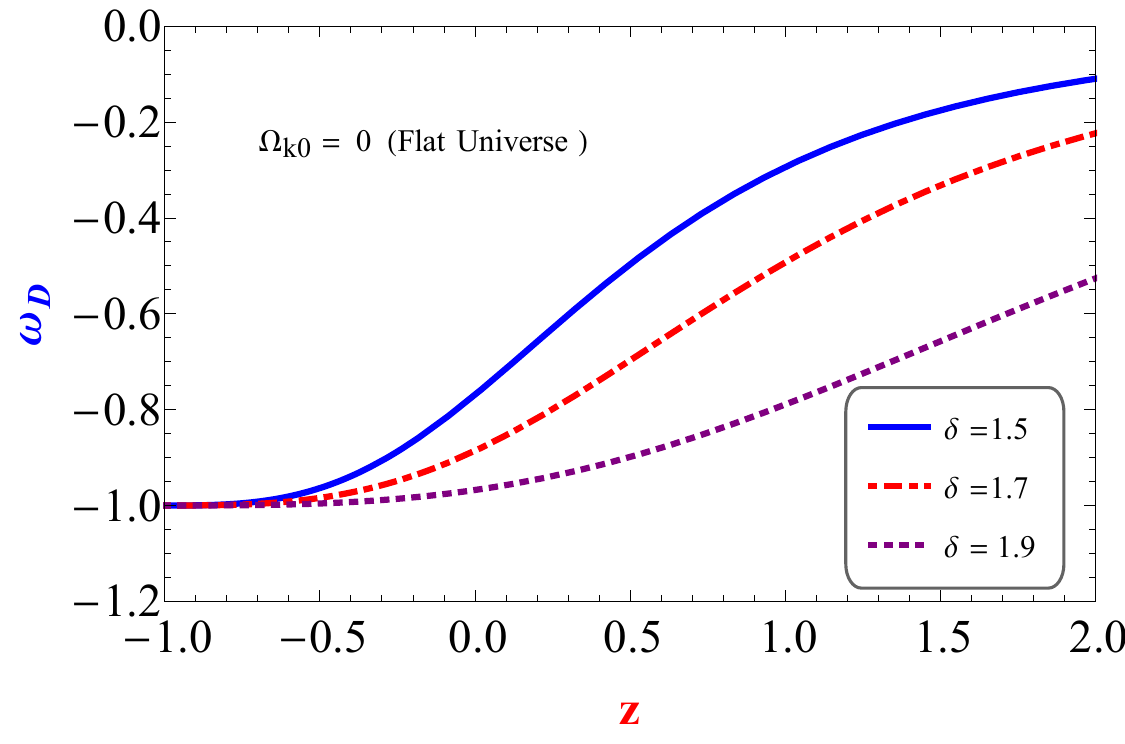}
		\caption {The evolution of equation of state parameter  ($\Omega_{D}$)  in THDE model against redshift $z$ for different cases of model parameter $\delta = 1.5$, $\delta = 1.7$ and $\delta = 1.9$. Selected graphs are plotted for $\Omega_{D0}= 0.72$ and  taking $\Omega_{K0}$= $ -0.0027$, $0.001$ and  $0$  corresponding to open, closed and flat  universes, respectively, in the light of $WMAP + eCMB + BAO + H_{0}$ observational data. } 
		
	\end{center}
\end{figure}

\section{ Cosmological behaviour of the non - interactiing THDE model}
Substituting  Eq. (\ref{eq13}) in Eq. (\ref{eq9}), and combining with Eq. (\ref{eq12}), we obtained
\begin{eqnarray}
\label{eq15}
\omega _D=-\frac{(\delta -1) \left(\Omega _K+1\right)}{(\delta -2) \Omega _D+\Omega _K+1}
\end{eqnarray}

and
\begin{eqnarray}
\label{eq16}
q=\frac{\left(\Omega _K+1\right) \left((1-2 \delta ) \Omega _D+\Omega _K+1\right)}{2 \left((\delta -2) \Omega _D+\Omega _K+1\right)}
\end{eqnarray}
In order to describe cosmic evolution, we have obtained cosmological parameters, deceleration parameter $q$ and THDE EoS parameter $\omega_{D}$ which is given by Eqs. (\ref{eq15}) and (\ref{eq16}). In Fig. 1, we have plotted the evolutionary behaviour of the deceleration parameter $q$ versus redshift $z$ for different Tsallis model parameter $\delta$ and also different contributions of the spatial curvature. From this graph, the transition from decelerated to the accelerated universe occurs from closed, flat and open universes. Moreover, the difference between them is minor. our results show that the transition redshift from deceleration to accelerated phase lies in the interval $0.4 <z<0.6$, which is fully consistent with the observations \cite{ref101}.\\

The behaviour of THDE EoS parameter $\omega_{D}$ against redshift $z$ for
THDE with the apparent horizon cutoff has been plotted in Fig.
2, for different Tsallis model parameter $\delta$ and also different contributions of the spatial curvature. From this figure, we may clearly see that the THDE model with the apparent cutoff can lead to
accelerated expansion, and $\omega_{D}$
tends to $-1$ at the future for closed, flat and open universes which implies that this model
mimics the cosmological constant at late time.

\section{Statefinder analysis for non - interacting THDE}

Now, we discuss the statefinder analysis of the THDE
model. It ought to be referenced that the statefinder analysis with $\omega_{D}-\omega_{D}'$ pair for THDE in the flat universe has been examined in detail in \cite{ref70,ref71}, where the attention is put on the analysis of the various interims of parameter $\delta$. In \cite{ref70,ref71}, it has been exhibited that
from the statefinder perspective, $\delta$ assumes a key role. Here we need to concentrate on the statefinder analysis of the contributions of the spatial curvature. It is important that in the non-flat universe the $\Lambda$CDM model does not compare a fixed point in the statefinder plane, it displays an evolution trajectory as\\

\begin{eqnarray}
\Big(s,r\Big)_{nonflat-\Lambda CDM} = \Big(0,\Omega_{total}\Big)
\end{eqnarray}

\begin{figure}
	\begin{center}
		\includegraphics[width=16cm,height=7cm, angle=0]{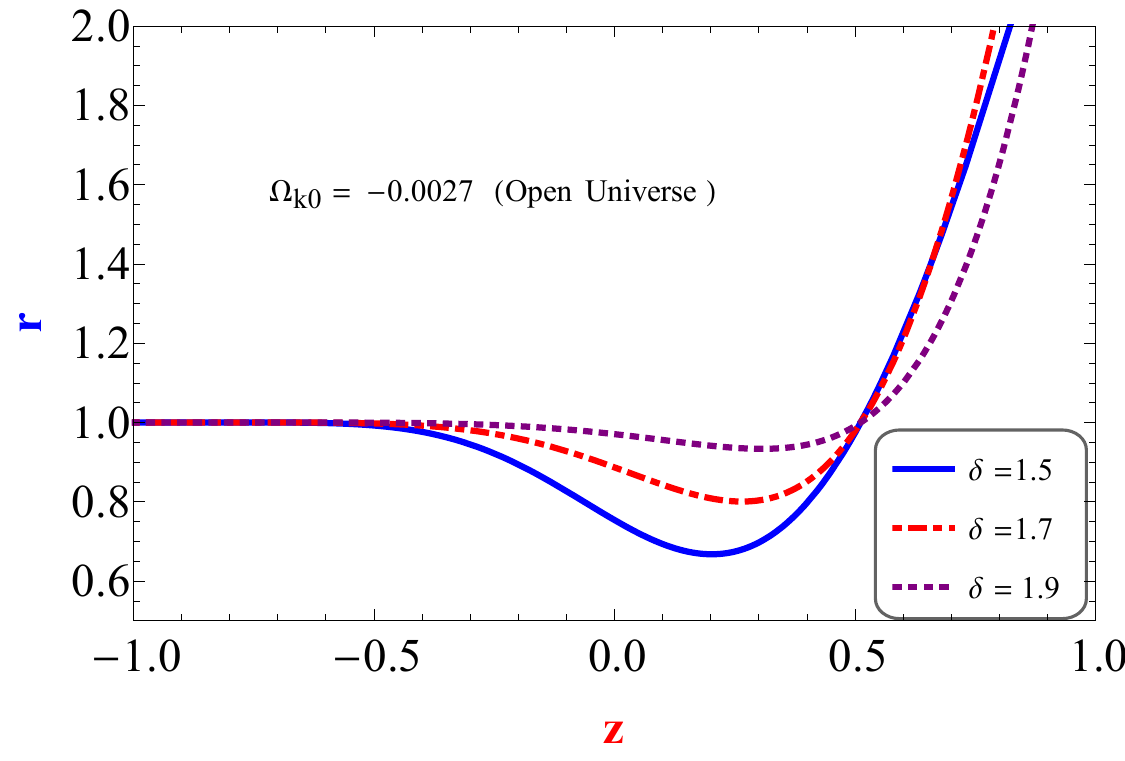}
		\includegraphics[width=16cm,height=7cm, angle=0]{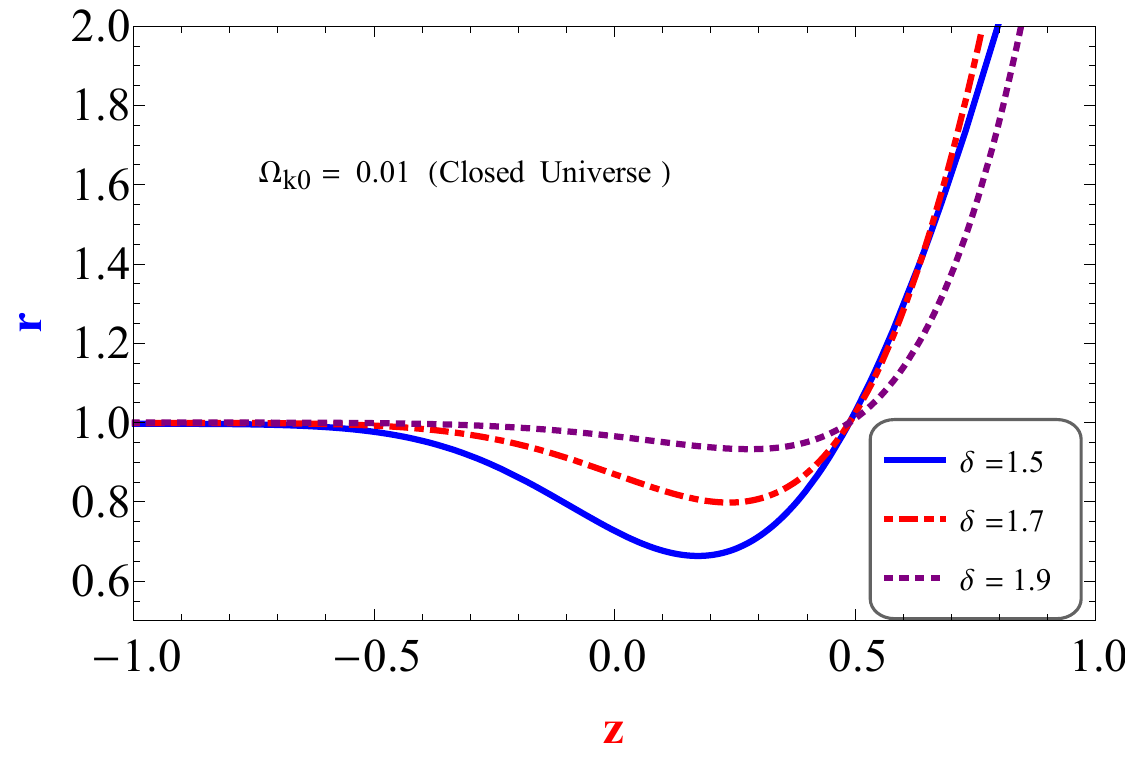}
		\includegraphics[width=16cm,height=7cm, angle=0]{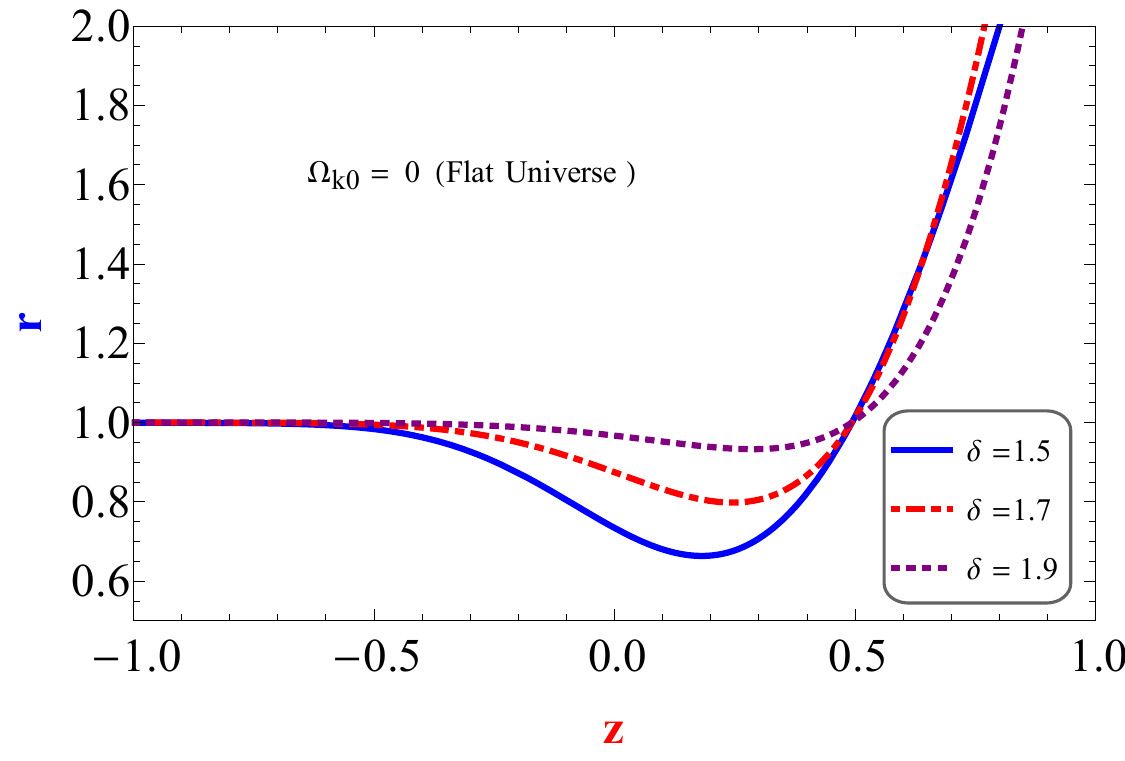}	
			\caption {The evolutionary behaviour  of first statefinder parameter  ($r$)  in THDE model against redshift $z$ for different cases of Tsallis parameter $\delta = 1.5$, $\delta = 1.7$ and $\delta = 1.9$. Selected graphs are plotted for $\Omega_{D0}= 0.72$ and taking $\Omega_{K0}$= $ -0.0027$, $0.001$ and  $0$  corresponding to open, closed and flat  universes, in the light of $WMAP + eCMB + BAO + H_{0}$ observational data.}
	\end{center}
\end{figure}

\begin{figure}
	\begin{center}
		\includegraphics[width=16cm,height=7cm, angle=0]{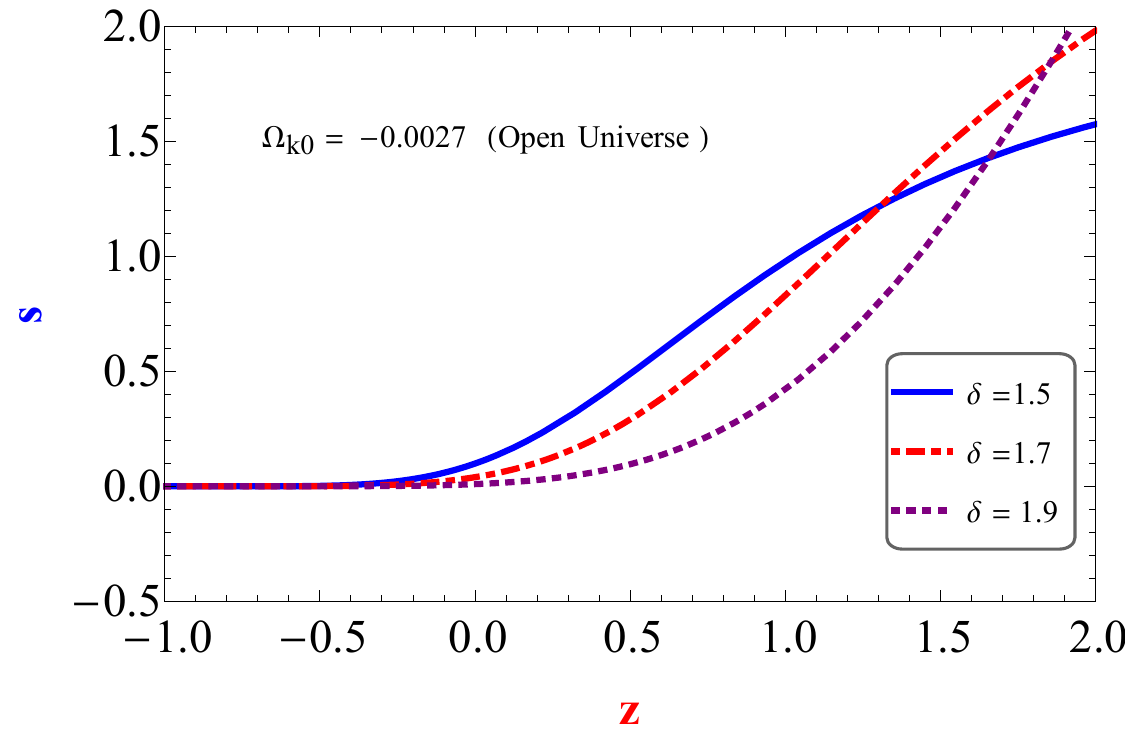}
		\includegraphics[width=16cm,height=7cm, angle=0]{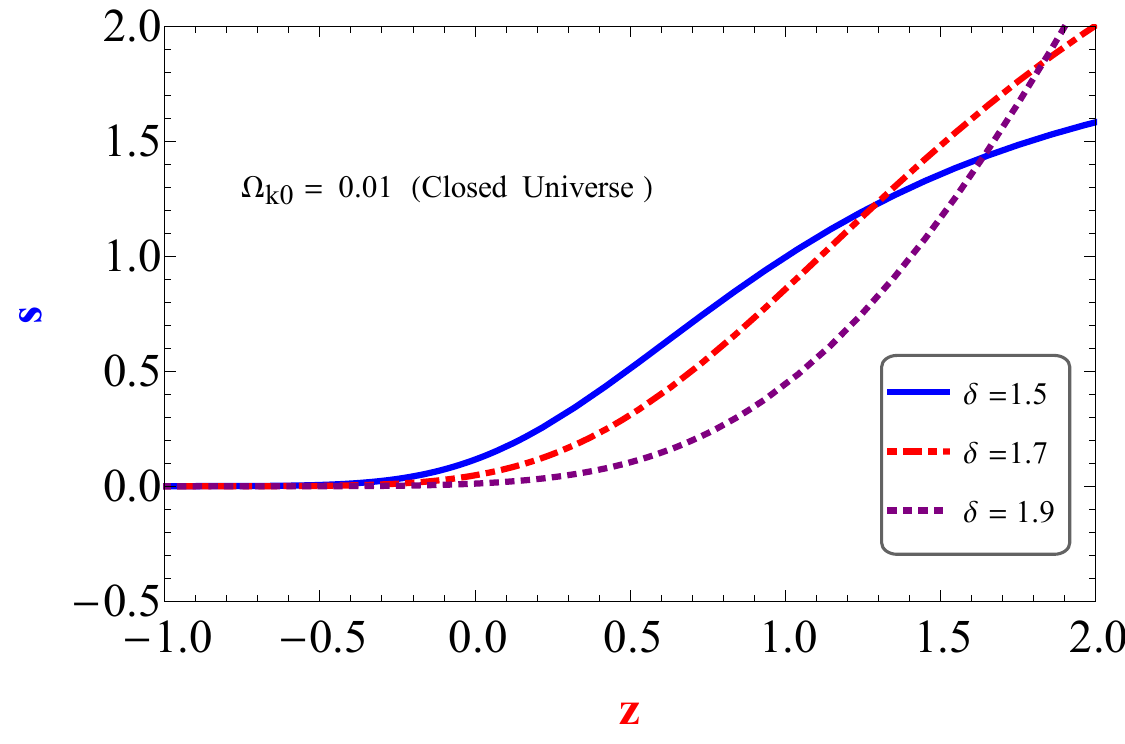}
		\includegraphics[width=16cm,height=7cm, angle=0]{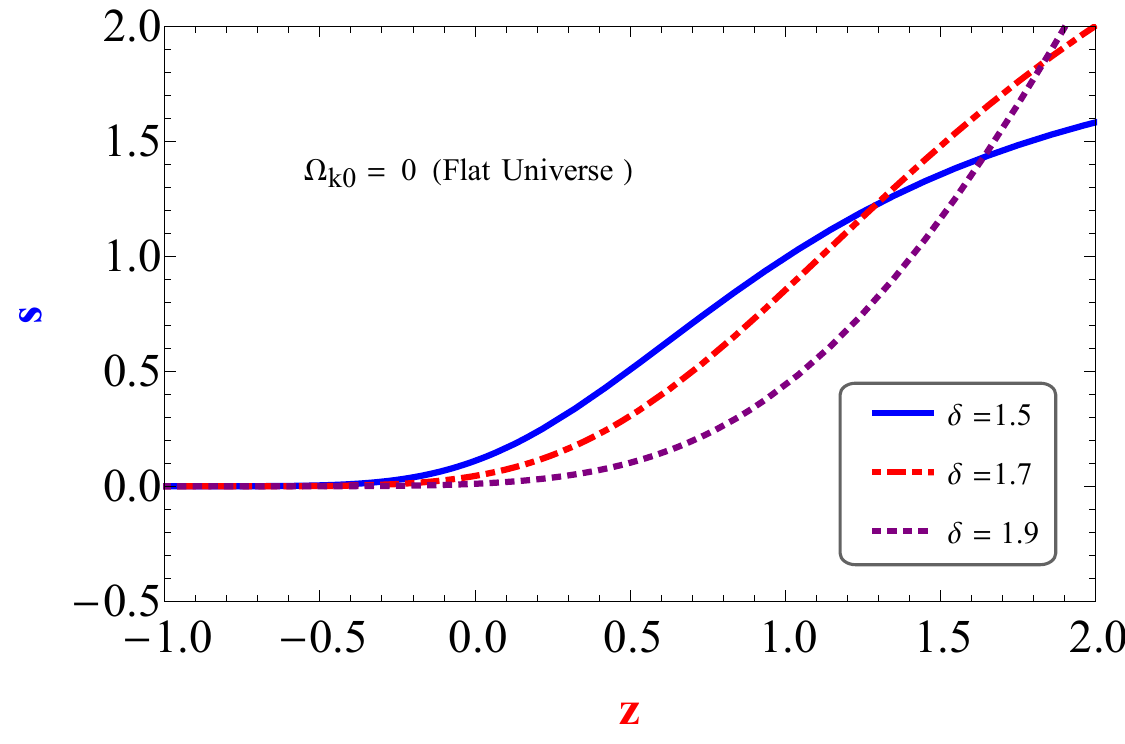}
		\caption {The evolutionary behaviour  of the second statefinder parameter  ($s$)  in THDE model against redshift $z$ for different cases of Tsallis parameter $\delta = 1.5$, $\delta = 1.7$ and $\delta = 1.9$. Selected graphs are plotted for $\Omega_{D0}= 0.72$ and taking $\Omega_{K0}$= $ -0.0027$, $0.001$ and  $0$  corresponding to open, closed and flat  universes, respectively, in the light of $WMAP + eCMB + BAO + H_{0}$ observational data.}
	\end{center}
\end{figure}

The statefinder parameters $r$ and $s$ may also be expressed in terms of EoS parameter and energy density as follows \cite{ref61} :
\begin{eqnarray}
\label{eq20}
r= \Omega_{total}+ \frac{9}{2}\omega_{D}(1+\omega_{D})\Omega_{D}-\frac{3}{2}\omega_{D}^{'}\Omega_{D}
\end{eqnarray}
\begin{eqnarray}
\label{eq21}
s= 1+ \omega_{D}-\frac{1}{3}\frac{\omega_{D}^{'}}{\omega_{D}}
\end{eqnarray}
 Note that the parameter $r$
is also called cosmic jerk and $ \Omega_{total}= 1+  \Omega_{K} =  \Omega_{m}+ \Omega_{D}$, is the total energy density.\\

 The statefinder parameter are obtained for THDE model as :
\begin{eqnarray}
\label{eq23}
r= \frac{9 (\delta -2) (\delta -1) \Omega _D \left(\Omega _K+1\right) \left(-\Omega _D+\Omega _K+1\right){}^2}{2 \left((\delta -2) \Omega _D+\Omega _K+1\right){}^3}+1
\end{eqnarray}

\begin{eqnarray}
\label{eq24}
s= -\frac{(\delta -2) \left(-\Omega _D+\Omega _K+1\right){}^2}{\left((\delta -2) \Omega _D+\Omega _K+1\right){}^2}
\end{eqnarray}

The first and second statefinder parameters against redshift have been plotted in Fig. 3 and 4. From these figures, we see that, both, first and second statefinder parameter at low red-shift of THDE approaches that of $\Lambda$CDM i.e. $r=1$ and $s=0$, for different Tsallis model parameter $\delta$ and also open, flat and closed universes while
at large $\Lambda$CDM deviates significantly from the standard behaviour. For Oscillating dark energy (ODE), the second statefinder parameter
at low red-shift approaches that of $\Lambda$CDM, while
at large $\Lambda$CDM deviates significantly from the standard
behaviour. The opposite holds for the first parameter \cite{ref65}. Cao et al \cite{ref64}, demonstrates the evolution of $r(z)$ and $s(z)$ versus redshift $z$ for $f(R)$ theory. They have shown different features for these parameters. Cui and Zhang \cite {ref102}, compared the HDE, NHDE, RDE, NADE and $\Lambda$CDM models with the first statefinder parameter $r(z)$ and differentiate them in the low-shift region. Thus, the THDE models differ from ODE and other dark energy models in terms of the first and second statefinder parameter. It is of interest to see that the cases are distinctively differentiated in the low - redshift region.\\

\begin{figure}
	\begin{center}
			\includegraphics[width=16cm,height=7cm, angle=0]{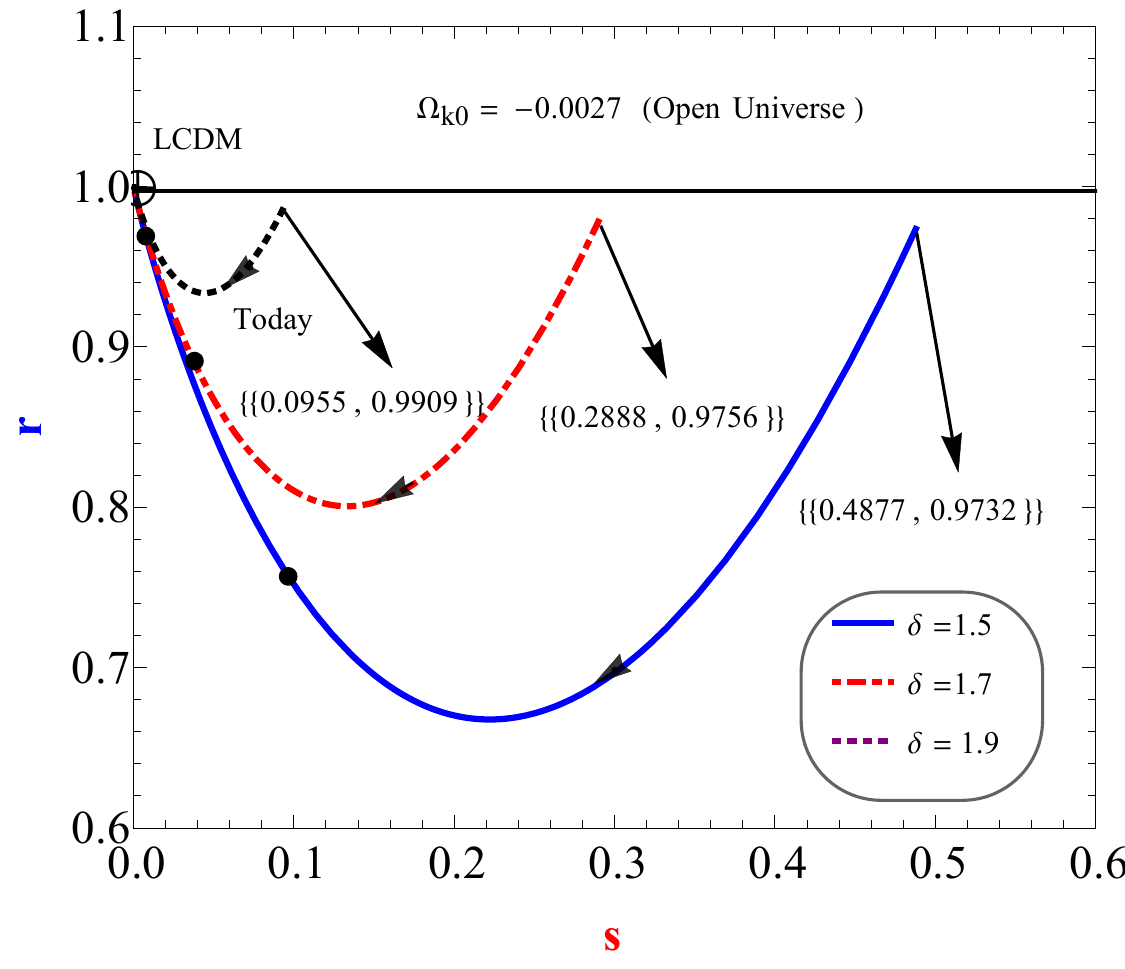}
			\includegraphics[width=16cm,height=7cm, angle=0]{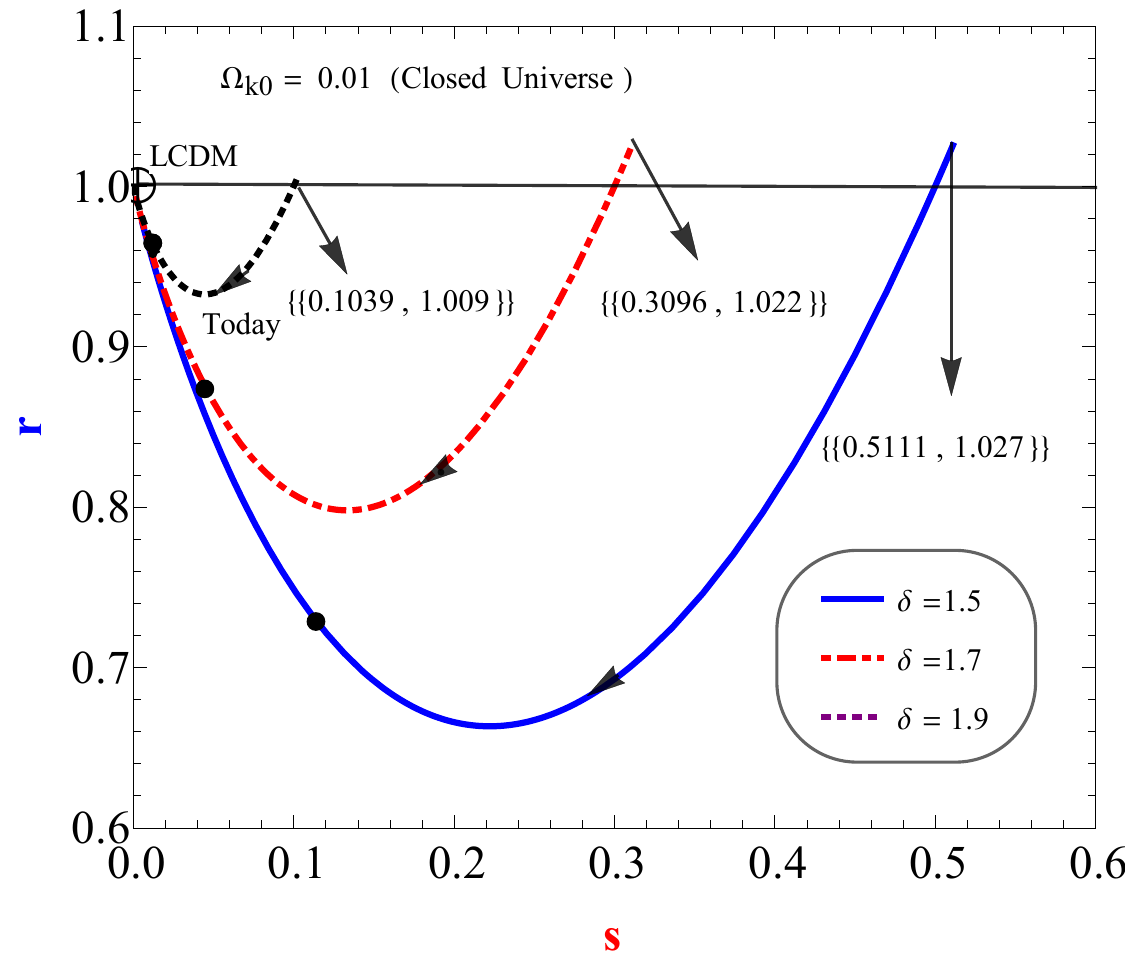}
			\includegraphics[width=16cm,height=7cm, angle=0]{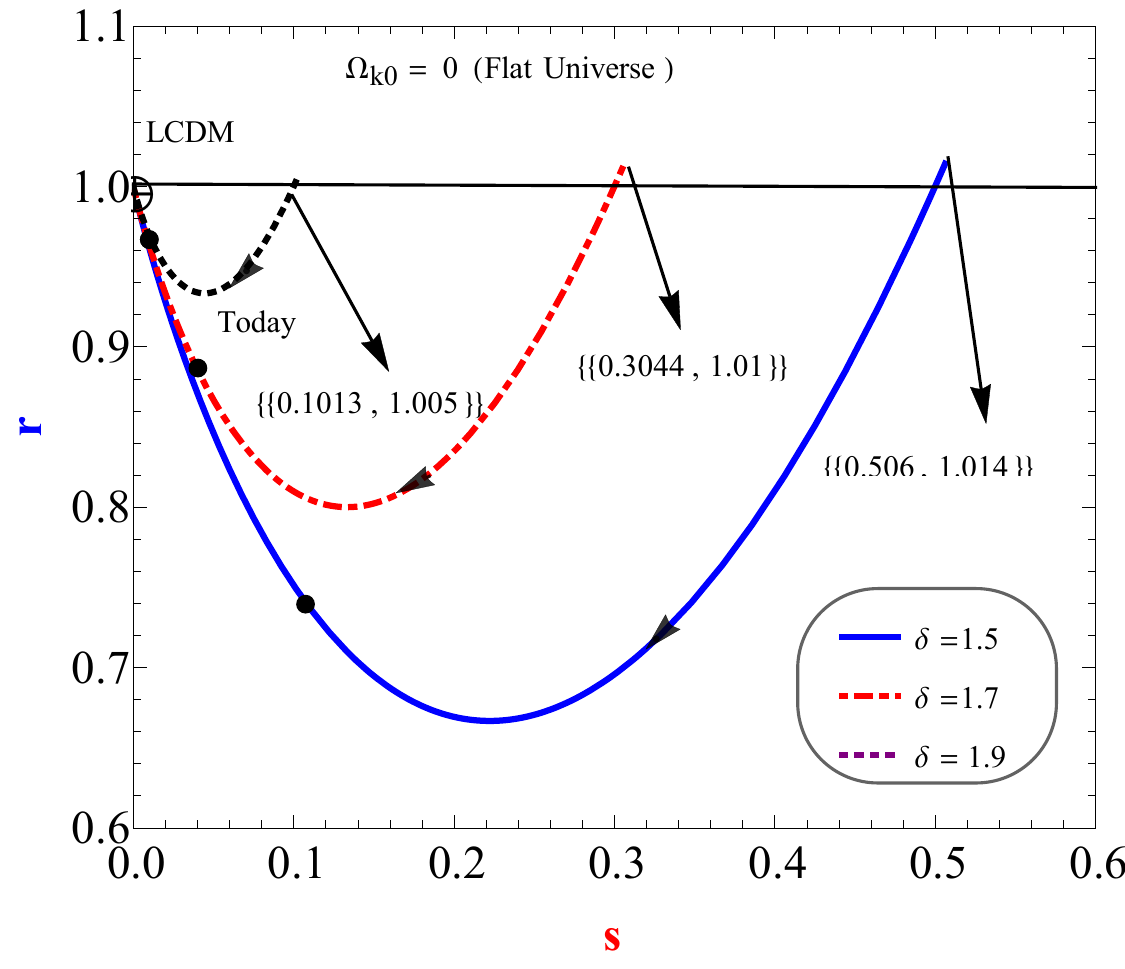}
		\caption {The  evolutionary trajectory in the  $s-r$ plane in THDE model for different cases of Tsallis parameter $\delta = 1.5$, $\delta = 1.7$ and $\delta = 1.9$. Selected graphs are plotted for $\Omega_{D0}= 0.72$ and  taking $\Omega_{K0}$= $ -0.0027$, $0.001$ and  $0$  corresponding to open, closed and flat  universes, respectively, in the light of $WMAP + eCMB + BAO + H_{0}$ observational data. LCDM corresponds the fixed point $(0, 1)$  which is shown by circle with plus sign in the figure. The present values of ($s_{0}, r_{0}$) are represented by solid- dots circles.}
	\end{center}
\end{figure}

\begin{figure}
	\begin{center}
		\includegraphics[width=16cm,height=7cm, angle=0]{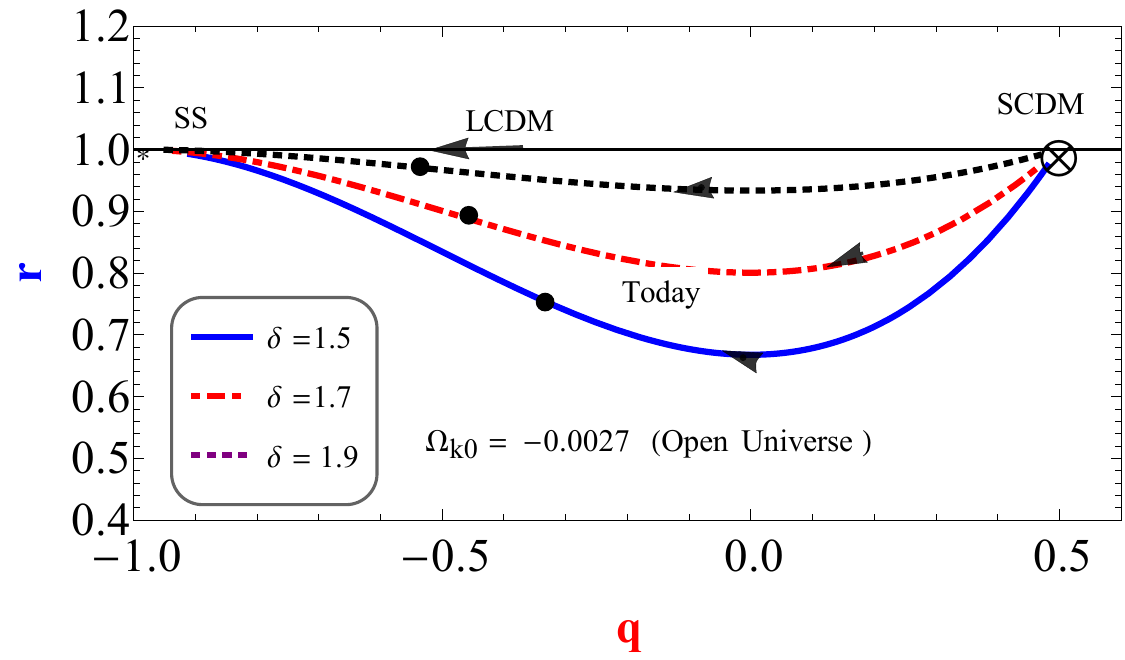}
		\includegraphics[width=16cm,height=7cm, angle=0]{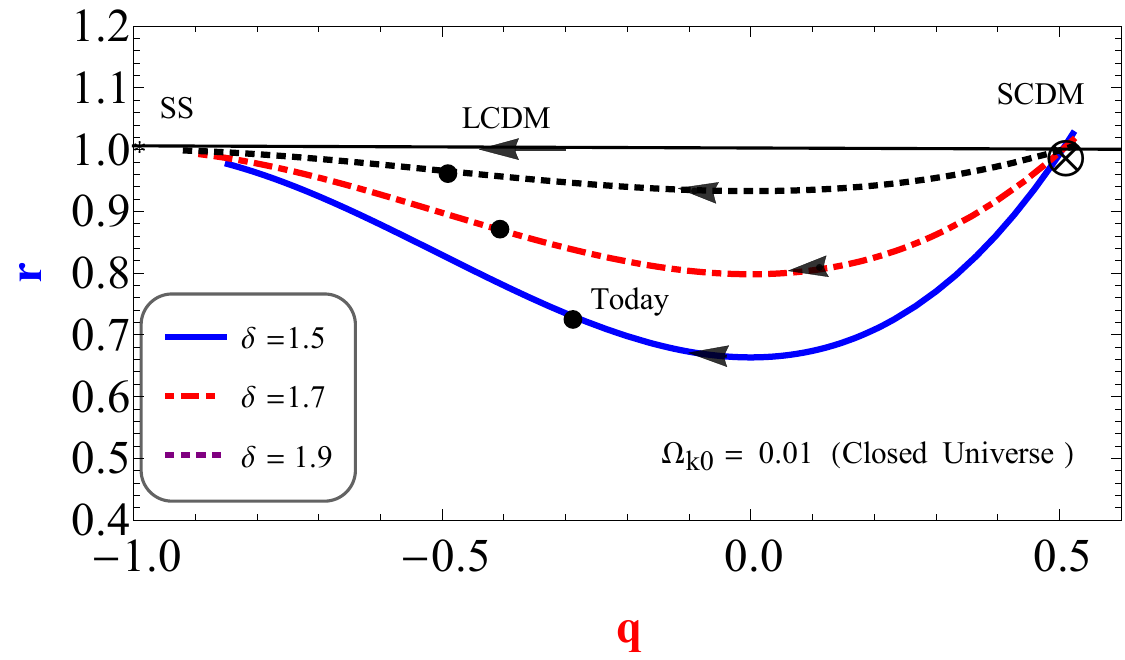}
		\includegraphics[width=16cm,height=7cm, angle=0]{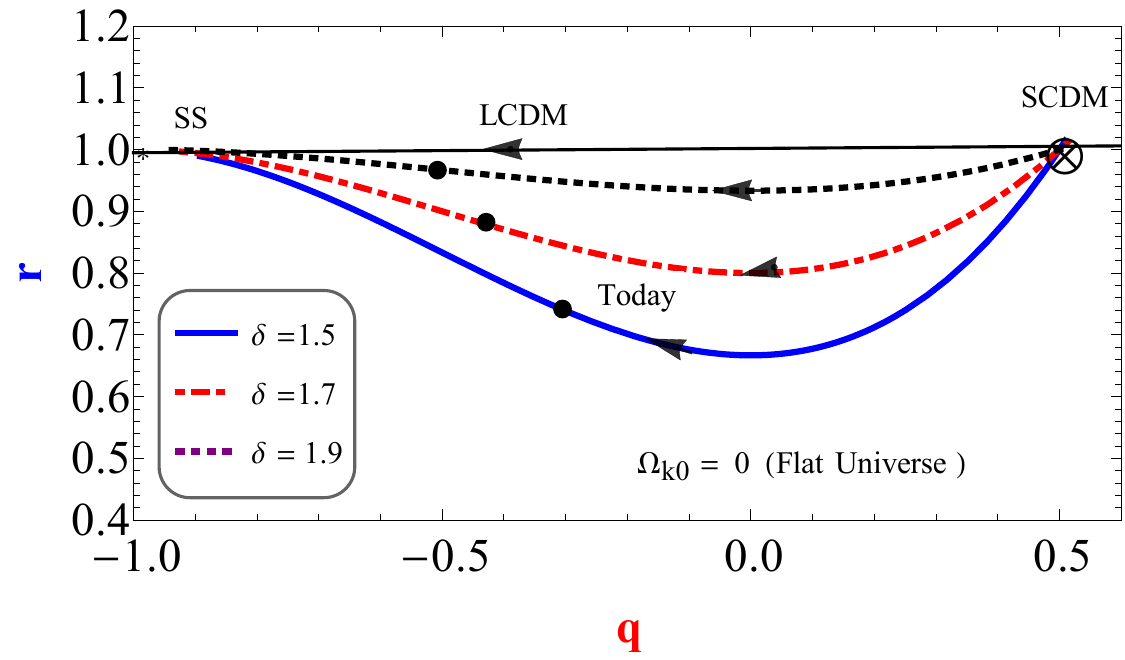}	
	\caption {The evolutionary trajectory in the  $q-r$ plane in THDE model for different cases of Tsallis parameter $\delta = 1.5$, $\delta = 1.7$ and $\delta = 1.9$ and for different spatial curvature, in the light of $WMAP + eCMB + BAO + H_{0}$ observational data. The de - sitter expansion - the steady state (SS)-  shown by star symbols is the fixed point $(-1, 1)$, and $(0.5, 1)$ denotes the SCDM (matter dominated)universe  shown by circle with cross sign.}
	\end{center}
\end{figure}
We have plotted the evolution trajectories in the statefinder $(r, s)$ and $(r, q )$
planes for our THDE model in Fig. 5 and Fig. 6, for different Tsallis parameter $\delta$ and also different contributions of the spatial curvature as
$\Omega_{K0}$ as -0.0027, 0 and 0.01
corresponding to the open, flat and closed universes, respectively. From Fig. 5, in $(r, s)$ evolutionary plane, we see that the derived THDE model, for open universe starts its evolutionary trajectories from $s = 0.4877$, $r = 0.9732$ for $\delta=1.5$, $s =0.2888 $, $r =0.9756 $ for $\delta=1.7$ and $s =0.0955 $, $r =0.9909 $ for $\delta=1.9$, for closed universe start its evolution from $s = 0.5111$, $r =1.027 $ for $\delta=1.5$, $s = 0.3096$, $r =1.022 $ for $\delta=1.7$ and $s = 0.1039$, $r =1.009 $ for $\delta=1.9$ and for flat universe start its evolution from $s = 0.505$, $r = 1.014$ for $\delta=1.5$, $s =0.3044 $, $r =1.01 $ for $\delta=1.7$ and $s =0.1013 $, $r =1.005 $ for $\delta=1.9$ but ends at
the $LCDM$ ( $r = 1$, $s = 0$) in the future for the open, flat and closed universes. One can also observe from this figure that the Tsallis HDE model would tends to the flat $LCDM$ model in the future for the open, flat and closed universes. The ($r-s$) plane behaviour of Tsallis HDE models differs from \cite{ref63a}. \\

From Fig. 6, in $(r, q)$ evolutionary plane, we observe that the Tsallis HDE model evolutionary trajectories started from matter dominated universe i.e. SCDM ( $r = 1$, $q = 0.5$) in the past, and their evolutionary trajectories approaches the
point ($q = -1$, $r = 1$) in the future i.e. the de Sitter expansion ($SS$) for the open, flat and closed universes. Moreover, the difference between them is minor, but we can see that, from Fig. 6, the distance from the de Sitter expansion for $\delta= 1.9$ is less in an open universe in comparison to closed and flat universes. Setare et al. \cite{ref61}, have a diagnosis the HDE in the non-flat universe with future event horizon as IR cutoff where the focus was different values of parameter $c$ and the spatial curvature contributions. The ($r- q$) plane behaviour of THDE models also differs from \cite{ref61}.\\

These graphs demonstrate that the evolutionary trajectories with distinct $\delta$ display altogether various highlights in the statefinder plane, which has been talked about in detail in \cite{ref70,ref71}. Presently we are keen on the analysis to the spatial curvature contributions in the derived models utilizing the statefinder parameters as a test. It has unmistakably appeared in Fig 5. also, Fig. 6, that the statefinder analysis has the intensity of testing the contributions of spatial curvature.\\

	\begin{figure}
	\begin{center}
		\includegraphics[width=16cm,height=7cm, angle=0]{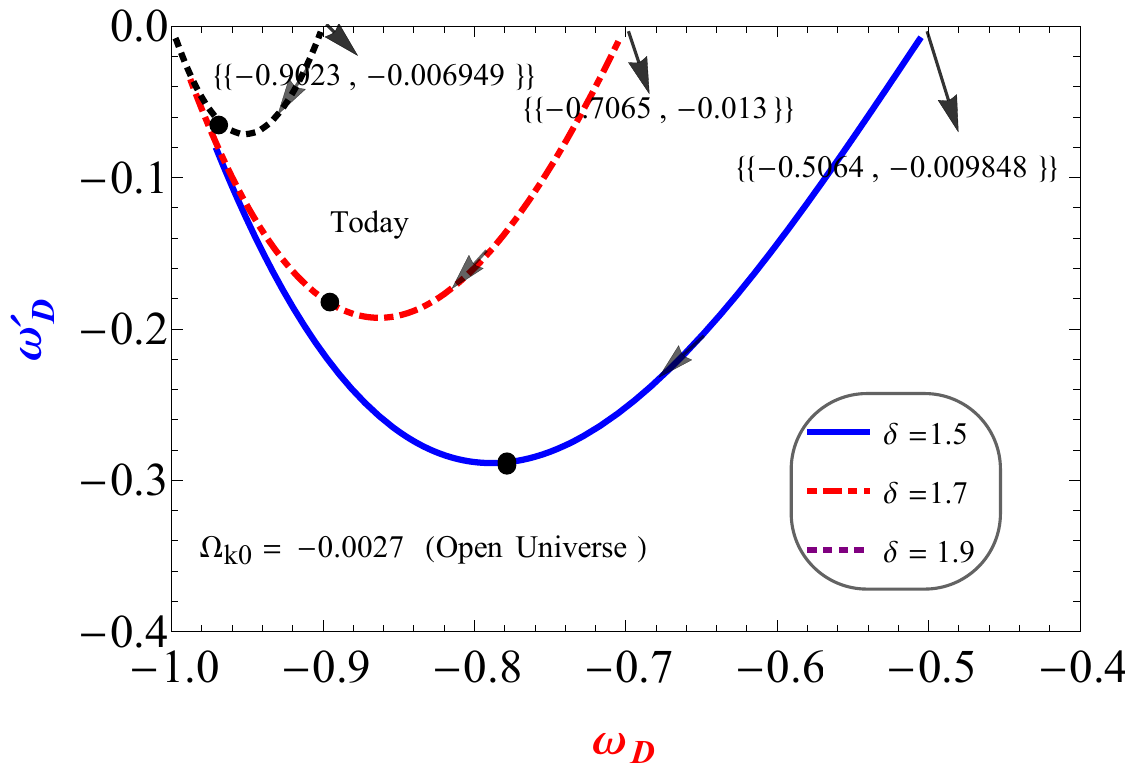}
		\includegraphics[width=16cm,height=7cm, angle=0]{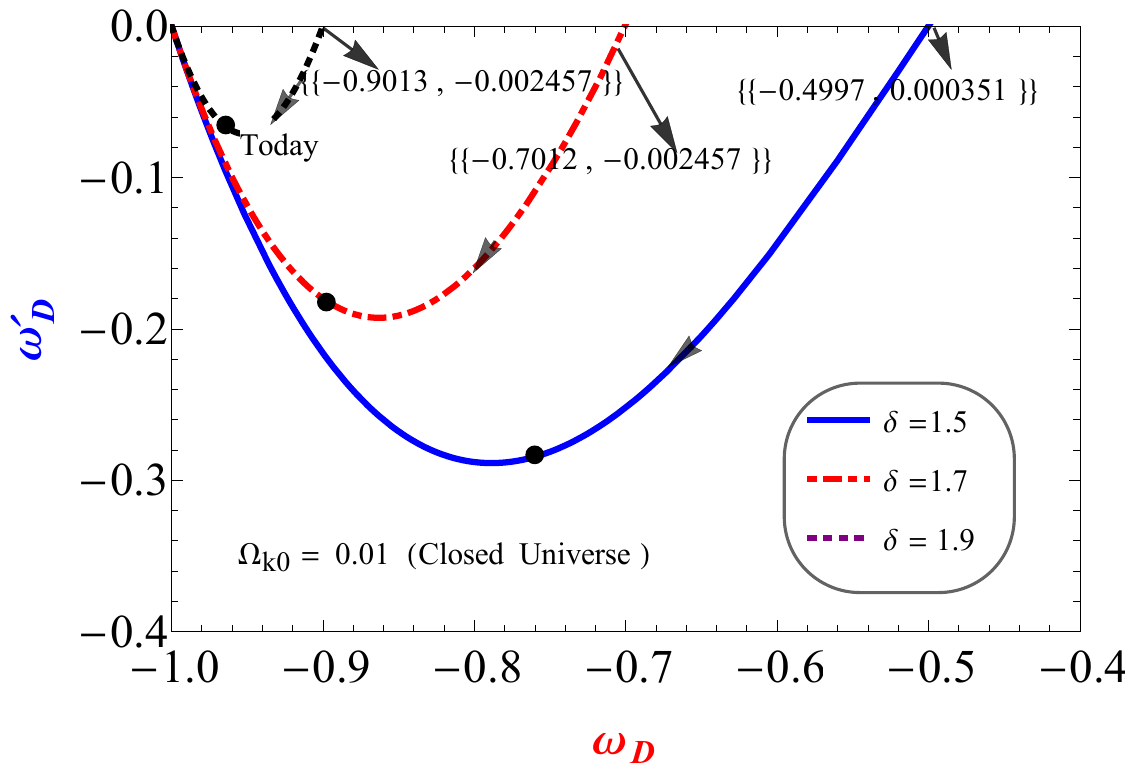}
		\includegraphics[width=16cm,height=7cm, angle=0]{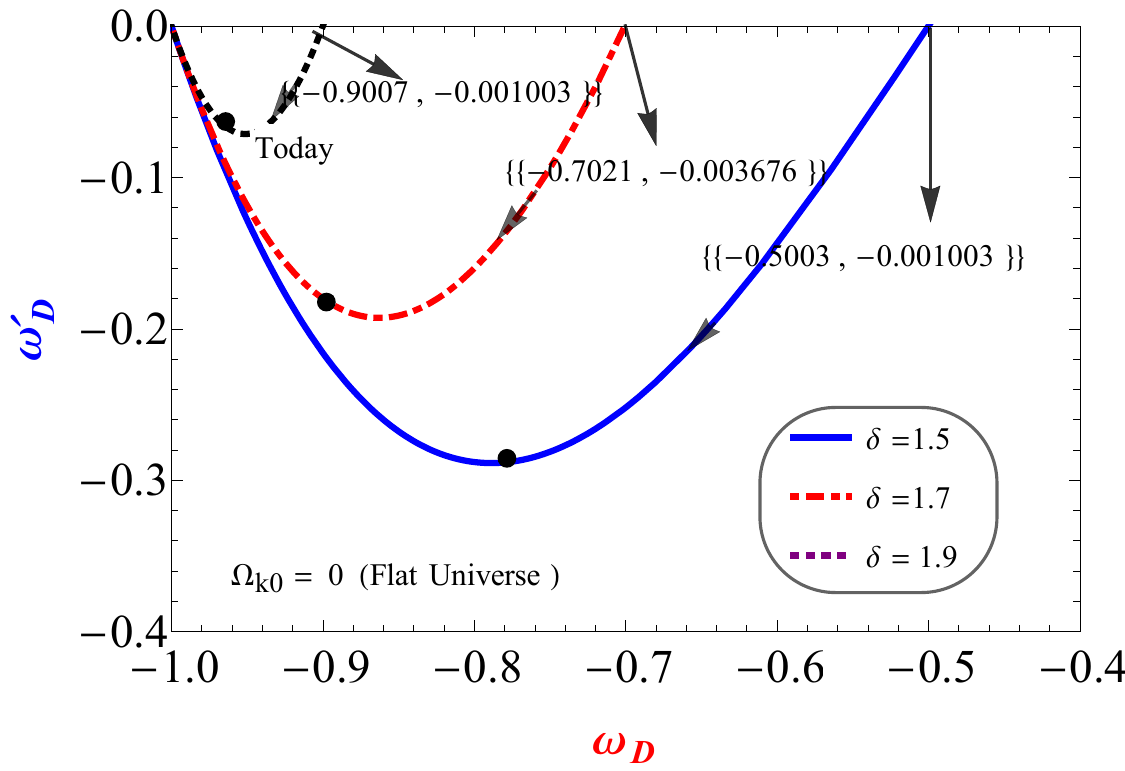}	
	\caption { The THDE evolution trajectories in the $\omega_{D}- \omega_{D}^{'}$  plane in THDE model for different cases of Tsallis parameter $\delta = 1.5$, $\delta = 1.7$ and $\delta = 1.9$. Selected graphs are plotted for $\Omega_{D0}= 0.72$ and taking $\Omega_{K0}$= $ -0.0027$, $0.001$ and  $0$  corresponding to open, closed and flat universes, respectively, in the light of $WMAP + eCMB + BAO + H_{0}$ observational data. The present values of ($\omega_{D},  \omega_{D}^{'}$ ) is denoted by the  solid dots circle.}
	\end{center}
\end{figure}
	\begin{table}
	\caption{\small The present values of the paramters $r, s, q, \omega_{D}$ and $\omega_{D}^{'}$ for different Tsallis parameter $\delta$ in different spatial curvature contributions.}
	\setlength{\arrayrulewidth}{0.1mm}
	\setlength{\tabcolsep}{0.1pt}
	\renewcommand{\arraystretch}{2.5}
	
	\begin{center}
		
		\begin{tabular}{|c|c|c|c|c|c|c|c|c|c|}
			
			\hline
			Curvature  &\multicolumn{3}{|c|}{$\Omega_{k}$ =-0.0027 (open universe)}   & \multicolumn{3}{|c|}{$\Omega_{k}$ = 0.01 (Closed Universe) }&\multicolumn{3}{|c|}{ $\Omega_{k}$ = 0 ( Flat Universe)}  \\
			\hline
			\small Parameter&\small $\delta$ = 1.5&\small $\delta$ = 1.7&\small $\delta$ = 1.9&\small $\delta$ = 1.5&\small $\delta$ = 1.7&\small $\delta$ = 1.9&\small $\delta$ = 1.5& \small$\delta$ = 1.7&\small $\delta$ = 1.9\\
			\hline
			\small r &\small 0.75403&\small 0.886881&\small 0.970638& \small 0.727003& \small 0.869757& \small 0.965314& \small 0.741921& \small0.879249&0.968279\\
			\hline
			\small s &\small 0.985806&\small 0.0395844&\small 0.00944539&\small 0.115778&\small 0.047642&\small 0.0115643&\small 0.106509&\small 0.0432623&\small 0.0104058\\
			\hline
			\small q &\small -0.333056 &\small -0.453901 &\small -0.537545&\small -0.280977&\small -0.406264&\small -0.494803&\small -0.307692&\small -0.43038&\small -0.516129\\
			\hline
			\small  $ \omega _D $ &\small -0.777986 &\small -0.891026 &\small -0.969267 &\small -0.759398 &\small -0.880448 &\small -0.965994 &\small -0.769231 &\small -0.886076 &\small -0.967742\\
			\hline
			\small $ \omega _D'$ &\small -0.288089 &\small -0.185484 &\small -0.0619011 &\small -0.284372 &\small -0.189938  &\small -0.0650366 &\small -0.286755 &\small -0.187835 &\small -0.063442\\ \hline
			
		\end{tabular}
	\end{center}
\end{table}

\subsection{Analysis of the $\omega_{D}- \omega_{D}^{'}$ pair}

In this section, the dynamical feature of the Tsallis HDE, as an enhancement to the statefinder analysis, in $\omega_{D}-\omega_{D}^{'}$ plane has investigated. This diagnostic received recognition to some extent for examining DE models. In the $\omega_{D}-\omega_{D}^{'}$ plane, the scopes of quintessence DE and its behaviour have researched in \cite{ref69}. Forwarding this point, by considering the dynamical property of other DE models including progressively more wide models was being utilized from the $\omega_{D}-\omega_{D}^{'}$ viewpoint \cite{ref103,ref104,ref105}, etc.. Recently the comparison among different DE models has been explored by the $\omega_{D}-\omega_{D}^{'}$ plane \cite{ref67}. The $\omega_{D}-\omega_{D}^{'}$ plane investigation provides us with an elective technique without uncertainty for classification of models of dark energy using the parameters depicting the dynamical property of DE. The estimation of the statefinder method is that the statefinder parameters are worked from the scale factor and its differential coefficient, and they are regularly to be evacuated in a model-self-governing way from observational data in spite of the way that it seems hard to achieve this at present. While the upside of the $\omega_{D}-\omega_{D}^{'}$ examination is that it is a direct dynamical investigation for DE. In this way, the statefinder $s - r$ is the geometrical end and the $\omega_{D}-\omega_{D}^{'}$ dynamical assurance can be viewed as complementarity in some sense.\\

Now, we will examine the THDE in the $\omega_{D}- \omega_{D}^{'}$ plane. Fig. 7 demonstrates an illustrative precedent in which we plot the transformative directions of the THDE model in the $\omega_{D}-\omega_{D}^{'}$ plane for various Tsallis parameter $\delta$ and furthermore various spatial curvature contributions, $\Omega_{K0}$ as - 0.0027, 0 and 0.01
comparing to the open, closed and flat universes, separately. The arrow in the plots represents the
evolutionary direction. We plainly observe that the parameter $\delta$ assumes an important role in the derived model:
the THDE behaves as quintessence- type dark energy with
$\omega_{D}\geq -1$, for all three values of parameter $\delta$ and different spatial curvature contributions. As appeared in these diagrams, the
value of $\omega_{D}$ decreases monotonically while the value of $\omega_{D}^{'}$ first decreases to a minimum then increases to zero. The effect of curvature contribution may be identified explicitly in this diagram, although the difference is minor. Henceforth, we see that the $\omega_{D} - \omega_{D}'$ dynamical conclusion may outfit with an important enhancement to the statefinder geometrical investigation.

\section{Conclusion}

We have explored the THDE in the nonflat universe within the framework of General Relativity filled with
matter and THDE considering apparent horizon as IR cutoff from the statefinder and $\omega_{D} - \omega_{D}'$ pair viewpoint. We have acquired different parameters of cosmology to comprehend the decelerated to the accelerated evolution of the universe in our described model. Main results are concluded as\\

$\bullet$ We inspect the DP of our model for different Tsallis parameter $\delta$ for distinct spatial curvature contributions and represents decelerated to accelerating phases of the universe depending on the parameter $\delta$ and also the type of spatial curvature, which is in good agreement with the current observations. \\

$\bullet$ In the proposed THDE model, the EoS parameter deviates in quintessence scenario and reaches to $\omega_{D} = -1$, at the future for different Tsallis parameter $\delta$ in different spatial curvature contributions. The model shows affirmity to $\Lambda$CDM model for all three values of $\delta$ in open, flat and closed universes.\\

$\bullet$ As suggested by various cosmological observations that the universe is in accelerated expansion phase presently, the numerous cosmological models including the modified gravity or DE have been proposed to translate this grandiose accelerated expansion. This prompts the issue of how to segregate between these different candidates. The statefinder analysis gives a helpful mechanism to break the conceivable depravity of various cosmological models by establishing the parameters ${r, s}$ utilizing the higher differential coefficient of the scale factor. In this way, the strategy for plotting the evolutionary trajectories of DE models in the statefinder plane can be utilized as a diagnostic tool to separate between various models.\\

Also, despite the fact that we are missing with regards to a fundamental hypothesis for the DE, this hypothesis is attempted to have a few highlights of a quantum gravity hypothesis, which can be investigated theoretically by considering the holographic rule of quantum gravity hypothesis. So THDE model gives us an endeavour to investigate the nature of DE inside a structure of the fundamental theory.\\

$\bullet$ $\omega_{D} - \omega_{D}'$ pair and the statefinder pair plots demonstrate that the spatial curvature contributions in the model can be analyzed expressly in this technique. \\

$\bullet$ The present values of the paramters $r, s, q, \omega_{D}$ and $\omega_{D}^{'}$ for different Tsallis parameter $\delta$ in different spatial curvature contributions are summarised in Table 2.\\

$\bullet$ The scenario of THDE guides to fascinating cosmological phenomenology. The universe shows the regular thermal history, in particular, the progressive order of the matter and DE period, with the change from deceleration to speeding up occurring, previously it results in a completely DE domination in future. \\

$\bullet$ Important results of our analysis is that the Tsallis HDE model is not quite the same as other DE models, for example, Holographic dark energy, Ricci dark energy from the statefinder perspective and furthermore not the same as these with $\Lambda HDE$ and NADE from the $\omega_{D} - \omega_{D}'$ pair perspective. \\

The composite null diagnostic and statefinder hierarchy are subjected to be examined in future to turn out to be all the more near the features of Tsallis HDE.

\section*{Acknowledgments}
The authors are appreciative to IUCAA, Pune, India for offering help and office to do this research work. The authors are also appreciative to A. Pradhan for his fruitful suggestions which improved the paper in present form.


\end{document}